\documentclass[12pt]{article}
\usepackage{euler}
\usepackage{amsmath}
\usepackage{amssymb}
\usepackage{amstext}
\usepackage{amscd}
\usepackage{amsfonts}
\DeclareMathAlphabet{\EuFrak}{U}{euf}{m}{n}
\DeclareMathAlphabet{\EuScript}{U}{eus}{m}{n}

\title{{\bf Superstring and Superstring Field Theory: a new
solution using Ultradistributions of Exponential Type}
\thanks{\it{This work was partially supported by Consejo
Nacional
de Investigaciones Cient\'{\i}ficas and Comisi\'{o}n de
Investigaciones Cient\'{\i}ficas de la Pcia. de Buenos
Aires;
Argentina.}}}
\author{C.G.Bollini and M.C.Rocca\\
Departamento de F\'{\i}sica, Fac. de Ciencias Exactas,\\
Universidad Nacional de La Plata.\\
C.C. 67 (1900) La Plata. Argentina.}

\date{August 17, 2008}

\begin{document}

\maketitle

\vspace{-5mm}

\begin{abstract}

In this paper we show that 
Ultradistributions of Exponential Type
(UET) are appropriate for the description
in a consistent way superstring and superstring field theories.
A new Lagrangian for the closed superstring is given.
We show that the superstring field 
is a linear superposition of UET of compact support,
and give the notion of anti-superstring.
We evaluate the propagator  for the superstring field,
and calculate the convolution of two of them.


\end{abstract}

\newpage

\setcounter{page}{2}

\renewcommand{\theequation}{\arabic{section}.\arabic{equation}}

\section{Introduction}

In a serie of papers \cite{tp1,tp2,tp3,tp4,tp5}
we have shown that Ultradistribution theory of 
Sebastiao e Silva  \cite{tp6,tp7,tp8} permits a significant advance in the treatment 
of quantum field theory. In particular, with the use of the 
convolution of Ultradistributions we have shown that it  is possible
to define a general product of distributions ( a product in a ring
with divisors of zero) that sheds new ligth on  the question of the divergences
in Quantum Field Theory. Furthermore, Ultradistributions of Exponential Type  
(UET) are  adequate for to describe
Gamow States and exponentially increasing fields in Quantum 
Field Theory \cite{tp9,tp10,tp11}.

Ultradistributions also have the
advantage of being representable by means of analytic functions.
So that, in general, they are easier to work with  and,
as we shall see, have interesting properties. One of those properties
is that Schwartz's tempered distributions are canonical and continuously
injected into  UET
and as a consequence the Rigged
Hilbert Space  with tempered distributions is  canonical and continuously
included
in the Rigged Hilbert Space with  Ultradistributions of 
Exponential Type.

Another interesting property is that the space of UET is 
reflexive under the operation of Fourier transform (in a way similar to that
of tempered distributions of Schwartz)

In a recent paper (\cite{tq1})  we have shown
that Ultradistributions of Exponential type
provides an adecuate framework 
for a consistent treatment of 
string and string field theories. In particular, a general
state of the closed bosonic bradyonic string is 
represented by UET of compact support,
and as a consequence the string field of a bradyonic 
bosonic string is a linear combination
of UET of compact support (CUET).

In this paper we extend the formalism developed in (\cite{tq1})
to the supersymmetric string.

This paper is organized as follows:
in sections 2 and 3 we define the Ultradistributions of Exponential Type
and their Fourier transform. They
are  part of a Guelfand's Triplet ( or Rigged Hilbert Space \cite{tp12} )
together with their respective duals and a ``middle term'' Hilbert
space. 
In sections 4 and 5 we give the main results obtained in (\cite{tq1})
to be applied in this paper.
In section 6 we treate the supersymmetric
string, giving a new lagrangian, defining the physical state 
of the string and solving the non-linear Euler-Lagrange
equations and the constraints. 
In section 7 we give a representation for the states 
of the string using CUET of compact support
Also we obtain the expression for a general state
of the  supersymmetric string.
In section 8 we give  expressions for the field of the string,
the string field propagator and the creation and anihilation
operators of a string. We define in a analog way to Quantum 
Field Theory the notion of anti-string and its corresponding
creation and anihilation fields.
In section 9, we give expressions for the non-local action of a free superstring
and a non-local interaction lagrangian for the string field inspired 
in Quantum Field Theory.
Also we show how to evaluate the convolution
of two superstring field propagators.
Finally, section 10 is reserved for a discussion of the principal results.

\section{Ultradistributions of Exponential Type}

Let ${\cal S}$ be the Schwartz space of rapidly decreasing test functions. 
Let ${\Lambda}_j$ be the region of the complex plane defined as:
\begin{equation}
\label{er2.1}
{\Lambda}_j=\left\{z\in\boldsymbol{\mathbb{C}} :
|\Im(z)|< j : j\in\boldsymbol{\mathbb{N}}\right\}
\end{equation}
According to ref.\cite{tp6,tp8} the space of test functions $\hat{\phi}\in
{\large{V}}_j$ is
constituted by all entire analytic functions of ${\cal S}$ for which
\begin{equation}
\label{ep2.2}
||\hat{\phi} ||_j=\max_{k\leq j}\left\{\sup_{z\in{\Lambda}_j}\left[e^{(j|\Re (z)|)}
|{\hat{\phi}}^{(k)}(z)|\right]\right\}
\end{equation}
is finite.\\
The space $\large{Z}$ is then defined as:
\begin{equation}
\label{er2.3}
\large{Z} =\bigcap_{j=0}^{\infty}{\large{V}}_j
\end{equation}
It is a complete countably normed space with the topology generated by
the system of semi-norms $\{||\cdot ||_j\}_{j\in \mathbb{N}}$.
The dual of ${\large{Z}}$, denoted by
${\large{B}}$, is by definition the space of ultradistributions of exponential
type (ref.\cite{tp6,tp8}).
Let ${S}$ be the space of rapidly decreasing sequences. According to
ref.\cite{tp12} ${S}$ is a nuclear space. We consider now the space of
sequences ${P}$ generated by the Taylor development of
$\hat{\phi}\in{\large{Z}}$
\begin{equation}
\label{er2.4}
{P}=\left\{{Q} : {Q}
\left(\hat{\phi}(0),{\hat{\phi}}^{'}(0),\frac {{\hat{\phi}}^{''}(0)} {2},...,
\frac {{\hat{\phi}}^{(n)}(0)} {n!},...\right) : \hat{\phi}\in{Z}\right\}
\end{equation}
The norms that define the topology of ${P}$ are given by:
\begin{equation}
\label{er2.5}
||\hat{\phi} ||^{'}_p=\sup_n \frac {n^p} {n} |{\hat{\phi}}^n(0)|
\end{equation}
${P}$ is a subespace of ${S}$ and therefore is a nuclear space.
As the norms $||\cdot ||_j$ and $||\cdot ||^{'}_p$ are equivalent, the correspondence
\begin{equation}
\label{er2.6}
{\large{Z}}\Longleftrightarrow {P}
\end{equation}
is an isomorphism and therefore ${Z}$ is a countably normed nuclear space.
We can define now the set of scalar products
\[<\hat{\phi}(z),\hat{\psi}(z)>_n=\sum\limits_{q=0}^n\int\limits_{-\infty}^{\infty}e^{2n|z|}
\overline{{\hat{\phi}}^{(q)}}(z){\hat{\psi}}^{(q)}(z)\;dz=\]
\begin{equation}
\label{er2.7}
\sum\limits_{q=0}^n\int\limits_{-\infty}^{\infty}e^{2n|x|}
\overline{{\hat{\phi}}^{(q)}}(x){\hat{\psi}}^{(q)}(x)\;dx
\end{equation}
This scalar product induces the norm
\begin{equation}
\label{er2.8}
||\hat{\phi}||_n^{''}=[<\hat{\phi}(x),\hat{\phi}(x)>_n]^{\frac {1} {2}}
\end{equation}
The norms $||\cdot ||_j$ and $||\cdot ||^{''}_n$ are equivalent, and therefore
${\large{Z}}$ is a countably hilbertian nuclear space.
Thus, if we call now ${{\large{Z}}}_p$ the completion of
${\large{Z}}$ by the norm $p$ given in (\ref{er2.8}), we have:
\begin{equation}
\label{er2.9}
{\large{Z}}=\bigcap_{p=0}^{\infty}{{\large{Z}}}_p
\end{equation}
where
\begin{equation}
\label{er2.10}
{{\large{Z}}}_0=\boldsymbol{H}
\end{equation}
is the Hilbert space of square integrable functions.\\
As a consequence the ``nested space''
\begin{equation}
\label{er2.11}
{\Large{U}}=\boldsymbol{(}{\large{Z}},
\boldsymbol{H}, {\large{B}}\boldsymbol{)}
\end{equation}
is a Guelfand's triplet (or a Rigged Hilbert space=RHS. See ref.\cite{tp12}).

Any Guelfand's triplet
${\Large{G}}=\boldsymbol{(}\boldsymbol{\Phi},
\boldsymbol{H},\boldsymbol{{\Phi}^{'}}\boldsymbol{)}$
has the fundamental property that a linear and symmetric operator
on $\boldsymbol{\Phi}$, admitting an extension to a self-adjoint
operator in
$\boldsymbol{H}$, has a complete set of generalized eigen-functions
in $\boldsymbol{{\Phi}^{'}}$ with real eigenvalues.

${\large{B}}$ can also be characterized in the following way
( refs.\cite{tp6},\cite{tp8} ): let ${{E}}_{\omega}$ be the space of
all functions $\hat{F}(z)$ such that:

${\Large {\boldsymbol{I}}}$-
$\hat{F}(z)$ is analytic for $\{z\in \boldsymbol{\mathbb{C}} :
|Im(z)|>p\}$.

${\Large {\boldsymbol{II}}}$-
$\hat{F}(z)e^{-p|\Re(z)|}/z^p$ is bounded continuous  in
$\{z\in \boldsymbol{\mathbb{C}} :|Im(z)|\geqq p\}$,
where $p=0,1,2,...$ depends on $\hat{F}(z)$.

Let ${N}$ be:
${N}=\{\hat{F}(z)\in{{E}}_{\omega} :\hat{F}(z)\; \rm{is\; entire\; analytic}\}$.
Then ${\large{B}}$ is the quotient space:

${\Large {\boldsymbol{III}}}$-
${\large{B}}={{E}}_{\omega}/{N}$

Due to these properties it is possible to represent any ultradistribution
as ( ref.\cite{tp6,tp8} ):
\begin{equation}
\label{er2.12}
\hat{F}(\hat{\phi})=<\hat{F}(z), \hat{\phi}(z)>=\oint\limits_{\Gamma} \hat{F}(z) \hat{\phi}(z)\;dz
\end{equation}
where the path ${\Gamma}$ runs parallel to the real axis from
$-\infty$ to $\infty$ for $Im(z)>\zeta$, $\zeta>p$ and back from
$\infty$ to $-\infty$ for $Im(z)<-\zeta$, $-\zeta<-p$.
( $\Gamma$ surrounds all the singularities of $\hat{F}(z)$ ).

Formula (\ref{er2.12}) will be our fundamental representation for a tempered
ultradistribution. Sometimes use will be made of ``Dirac formula''
for exponential ultradistributions ( ref.\cite{tp6} ):
\begin{equation}
\label{er2.13}
\hat{F}(z)\equiv\frac {1} {2\pi i}\int\limits_{-\infty}^{\infty}
\frac {\hat{f}(t)} {t-z}\;dt\equiv
\frac {\cosh(\lambda z)} {2\pi i}\int\limits_{-\infty}^{\infty}
\frac {\hat{f}(t)} {(t-z)\cosh(\lambda t)}\;dt
\end{equation}
where the ``density'' $\hat{f}(t)$ is such that
\begin{equation}
\label{er2.14}
\oint\limits_{\Gamma} \hat{F}(z) \hat{\phi}(z)\;dz =
\int\limits_{-\infty}^{\infty} \hat{f}(t) \hat{\phi}(t)\;dt
\end{equation}
(\ref{er2.13}) should be used carefully.
While $\hat{F}(z)$ is analytic on $\Gamma$, the density $\hat{f}(t)$ is in
general singular, so that the r.h.s. of (\ref{er2.14}) should be interpreted
in the sense of distribution theory.

Another important property of the analytic representation is the fact
that on $\Gamma$, $\hat{F}(z)$ is bounded by an exponential and a power of $z$
( ref.\cite{tp6,tp8} ):
\begin{equation}
\label{er2.15}
|\hat{F}(z)|\leq C|z|^pe^{p|\Re(z)|}
\end{equation}
where $C$ and $p$ depend on $\hat{F}$.

The representation (\ref{er2.12}) implies that the addition of any entire function
$\hat{G}(z)\in{N}$ to $\hat{F}(z)$ does not alter the ultradistribution:
\[\oint\limits_{\Gamma}\{\hat{F}(z)+\hat{G}(z)\}\hat{\phi}(z)\;dz=
\oint\limits_{\Gamma} \hat{F}(z)\hat{\phi}(z)\;dz+\oint\limits_{\Gamma}
\hat{G}(z)\hat{\phi}(z)\;dz\]
But:
\[\oint\limits_{\Gamma} \hat{G}(z)\hat{\phi}(z)\;dz=0\]
as $\hat{G}(z)\hat{\phi}(z)$ is entire analytic
( and rapidly decreasing ),
\begin{equation}
\label{er2.16}
\therefore \;\;\;\;\oint\limits_{\Gamma} \{\hat{F}(z)+\hat{G}(z)\}\hat{\phi}(z)\;dz=
\oint\limits_{\Gamma} \hat{F}(z)\hat{\phi}(z)\;dz
\end{equation}

Another very important property of ${\large{B}}$ is that
${\large{B}}$ is reflexive under the Fourier transform:
\begin{equation}
\label{er2.17}
{\large{B}}={\cal F}_c\left\{{\large{B}}\right\}=
{\cal F}\left\{{\large{B}}\right\}
\end{equation}
where the complex Fourier transform $F(k)$ of $\hat{F}(z)\in{\large{B}}$
is given by:
\[F(k)=\Theta[\Im(k)]\int\limits_{{\Gamma}_+}\hat{F}(z)e^{ikz}\;dz-
\Theta[-\Im(k)]\int\limits_{{\Gamma}_{-}}\hat{F}(z)e^{ikz}\;dz=\]
\begin{equation}
\label{er2.18}
\Theta[\Im(k)]\int\limits_0^{\infty}\hat{f}(x)e^{ikx}\;dx-
\Theta[-\Im(k)]\int\limits_{-\infty}^0\hat{f}(x) e^{ikx}\;dx
\end{equation}
Here ${\Gamma}_+$ is the part of $\Gamma$ with $\Re(z)>0$ and
${\Gamma}_{-}$ is the part of $\Gamma$ with $\Re(z)<0$
Using (\ref{er2.18}) we can interpret Dirac's formula as:
\begin{equation}
\label{er2.19}
F(k)\equiv\frac {1} {2\pi i}\int\limits_{-\infty}^{\infty}
\frac {f(s)} {s-k}\; ds\equiv{\cal F}_c\left\{{\cal F}^{-1}\left\{f(s)\right\}\right\}
\end{equation}
The treatment for ultradistributions of exponential type defined on
${\boldsymbol{\mathbb{C}}}^n$ is similar to the case of one variable.
Thus
\begin{equation}
\label{er2.20}
{\Lambda}_j=\left\{z=(z_1, z_2,...,z_n)\in{\boldsymbol{\mathbb{C}}}^n :
|\Im(z_k)|\leq j\;\;\;1\leq k\leq n\right\}
\end{equation}
\begin{equation}
\label{er2.21}
||\hat{\phi} ||_j=\max_{k\leq j}\left\{\sup_{z\in{\Lambda}_j}\left[
e^{j\left[\sum\limits_{p=1}^n|\Re(z_p)|\right]}\left| D^{(k)}\hat{\phi}(z)\right|\right]\right\}
\end{equation}
where $D^{(k)}={\partial}^{(k_1)}{\partial}^{(k_2)}\cdot\cdot\cdot{\partial}^{(k_n)}\;\;\;\;
k=k_1+k_2+\cdot\cdot\cdot+k_n$

${{\large{B}}}^n$ is characterized as follows. Let
${{E}}^n_{\omega}$ be the space of all functions $\hat{F}(z)$ such that:

${\Large {\boldsymbol{I}}}^{'}$-
$\hat{F}(z)$ is analytic for $\{z\in \boldsymbol{{\mathbb{C}}^n} :
|Im(z_1)|>p, |Im(z_2)|>p,...,|Im(z_n)|>p\}$.

${\Large {\boldsymbol{II}}}^{'}$-
$\hat{F}(z)e^{-\left[p\sum\limits_{j=1}^n|\Re(z_j)|\right]}/z^p$
is bounded continuous  in
$\{z\in \boldsymbol{{\mathbb{C}}^n} :|Im(z_1)|\geqq p,|Im(z_2)|\geqq p,
...,|Im(z_n)|\geqq p\}$,
where $p=0,1,2,...$ depends on $\hat{F}(z)$.

Let ${{N}}^n$ be:
${{N}}^n=\left\{\hat{F}(z)\in{{E}}^n_{\omega} :\hat{F}(z)\;\right.$
is entire analytic at minus in one of the variables $\left. z_j\;\;\;1\leq j\leq n\right\}$
Then ${{\large{B}}}^n$ is the quotient space:

${\Large {\boldsymbol{III}}}^{'}$-
${{\large{B}}}^n={{E}}^n_{\omega}/{{N}}^n$
We have now
\begin{equation}
\label{er2.22}
\hat{F}(\hat{\phi})=<\hat{F}(z), \hat{\phi}(z)>=\oint\limits_{\Gamma} \hat{F}(z) \hat{\phi}(z)\;
dz_1\;dz_2\cdot\cdot\cdot dz_n
\end{equation}
$\Gamma={\Gamma}_1\cup{\Gamma}_2\cup ...{\Gamma}_n$
where the path ${\Gamma}_j$ runs parallel to the real axis from
$-\infty$ to $\infty$ for $Im(z_j)>\zeta$, $\zeta>p$ and back from
$\infty$ to $-\infty$ for $Im(z_j)<-\zeta$, $-\zeta<-p$.
(Again $\Gamma$ surrounds all the singularities of $\hat{F}(z)$ ).
The n-dimensional Dirac's formula is
\begin{equation}
\label{ep2.23}
\hat{F}(z)=\frac {1} {(2\pi i)^n}\int\limits_{-\infty}^{\infty}
\frac {\hat{f}(t)} {(t_1-z_1)(t_2-z_2)...(t_n-z_n)}\;dt_1\;dt_2\cdot\cdot\cdot dt_n
\end{equation}
where the ``density'' $\hat{f}(t)$ is such that
\begin{equation}
\label{ep2.24}
\oint\limits_{\Gamma} \hat{F}(z)\hat{\phi}(z)\;dz_1\;dz_2\cdot\cdot\cdot dz_n =
\int\limits_{-\infty}^{\infty} f(t) \hat{\phi}(t)\;dt_1\;dt_2\cdot\cdot\cdot dt_n
\end{equation}
and the modulus of $\hat{F}(z)$ is bounded by
\begin{equation}
\label{er2.25}
|\hat{F}(z)|\leq C|z|^p e^{\left[p\sum\limits_{j=1}^n|\Re(z_j)|\right]}
\end{equation}
where $C$ and $p$ depend on $\hat{F}$.

\section{The Case N$\rightarrow\infty$}

\setcounter{equation}{0}

When the number of variables of the argument of the Ultradistribution of 
Exponential type tends to infinity we define:
\begin{equation}
\label{ep3.1}
d\mu(x)=\frac {e^{-x^2}} {\sqrt{\pi}}dx
\end{equation}
If $\hat{\phi}(x_1,x_2,...,x_n)$ is such that:
\begin{equation}
\label{ep3.2}
\idotsint\limits_{-\infty}^{\;\;\infty}|\hat{\phi}(x_1,x_2,...,x_n)|^2 d{\mu}_1d{\mu}_2...
d{\mu}_n<\infty
\end{equation}
where
\begin{equation}
\label{ep3.3}
d{\mu}_i=\frac {e^{-x_i^2}} {\sqrt{\pi}}dx_i
\end{equation}
Then by definition
$\hat{\phi}(x_1,x_2,...,x_n)\in L_2({\mathbb{R}}^n,\mu)$
and 
\begin{equation}
\label{ep3.4}
L_2({\mathbb{R}}^{\infty},\mu)=
\bigcup\limits_{n=1}^{\infty}L_2({\mathbb{R}}^n,\mu)
\end{equation}
Let $\hat{\psi}$ be given by
\begin{equation}
\label{ep3.5}
\hat{\psi}(z_1,z_2,...,z_n)={\pi}^{n/4}\hat{\phi}(z_1,z_2,...,z_n)
e^{\frac {z_1^2+z_2^2+...+z_n^2} {2}}
\end{equation}
where $\hat{\phi}\in {{\large{Z}}}^n$(the corresponding 
n-dimensional of ${\large{Z}}$).\\
Then by definition $\hat{\psi}(z_1,z_2,...,z_n)\in {\large{G}}({\mathbb{C}}^n)$,
\begin{equation}
\label{ep3.6}
{\large{G}}({\mathbb{C}}^{\infty})=\bigcup\limits_{n=1}^{\infty}
{\large{G}}({\mathbb{C}}^n)
\end{equation}
and the dual ${\large{G}}^{'}({\mathbb{C}}^{\infty})$ given by
\begin{equation}
\label{ep3.7}
{\large{G}}^{'}({\mathbb{C}}^{\infty})=\bigcup\limits_{n=1}^{\infty}
{\large{G}}^{'}({\mathbb{C}}^n)
\end{equation}
is the space of Ultradistributions of Exponential type.\\
The analog to (\ref{er2.11}) in the infinite dimensional case is:
\begin{equation}
\label{ep3.8}
{\Large{W}}=\boldsymbol{(}{\large{G}}({\mathbb{C}}^{\infty}),
L_2({\mathbb{R}}^{\infty},\mu), 
{\large{G}}^{'}({\mathbb{C}}^{\infty})\boldsymbol{)}
\end{equation}
If we define:
\begin{equation}
\label{ep3.9}
{\cal F}:{\large{G}}({\mathbb{C}}^{\infty})\rightarrow
{\large{G}}({\mathbb{C}}^{\infty})
\end{equation}
via the Fourier transform:
\begin{equation}
\label{ep3.10}
{\cal F}:{\large{G}}({\mathbb{C}}^n)\rightarrow
{\large{G}}({\mathbb{C}}^n)
\end{equation}
given by:
\begin{equation}
\label{ep3.11}
{\cal F}\{\hat{\psi}\}(k)=
\int\limits_{-\infty}^{\infty}\hat{\psi}(z_1,z_2,...,z_n)
e^{ik\cdot z+\frac {k^2} {2}}d{\rho}_1d{\rho}_2...d{\rho}_n
\end{equation}
where
\begin{equation}
\label{ep3.12}
d\rho(z)=\frac {e^{-\frac {z^2} {2}}} {\sqrt{2\pi}}\;dz
\end{equation}
we conclude that
\begin{equation}
\label{ep3.13}
{\large{G}}^{'}({\mathbb{C}}^{\infty})=
{\cal F}_c\{{\large{G}}^{'}({\mathbb{C}}^{\infty})\}=
{\cal F}\{{\large{G}}^{'}({\mathbb{C}}^{\infty})\}
\end{equation}
where in the one-dimensional case
\begin{equation}
\label{ep3.14}
{\cal F}_c\{\hat{\psi}\}(k)=
\Theta[\Im(k)]\int\limits_{{\Gamma}_+}\hat{\psi}(z)e^{ikz+\frac {k^2} {2}}\;d\rho-
\Theta[-\Im(k)]\int\limits_{{\Gamma}_{-}}\hat{\psi}(z)e^{ikz+\frac {k^2} {2}}\;d\rho
\end{equation}

\section{The Constraints for a Bradyonic Bosonic String}

\setcounter{equation}{0}

The constraints for a bradyonic bosonic string have been deduced
in ref.\cite{tq1}. 
We can describe  the bosonic string by a system composed of a
Lagrangian, one constraint and two initial conditions:
\begin{equation}
\label{ep4.1}
\begin{cases}
{\cal L}=|{\dot{X}}^2-{X^{'}}^2|\\
(\dot{X}+X^{'})^2=0\\
X_{\mu}(\tau,0)=X_{\mu}(\tau,\pi)=0
\end{cases}
\end{equation}
or equivalently
\begin{equation}
\label{ep4.2}
\begin{cases}
{\cal L}=|{\dot{X}}^2-{X^{'}}^2|\\
(\dot{X}-X^{'})^2=0\\
X_{\mu}(\tau,0)=X_{\mu}(\tau,\pi)=0
\end{cases}
\end{equation}

\section{A representation for  the states of the closed bosonic string}

\setcounter{equation}{0}

\subsection*{The case n finite}

From ref.\cite{tq1} we have
\begin{equation}
\label{ep5.1}
a=-z\;\;\;;\;\;\;a^+=\frac {d} {dz}
\end{equation}
and then
\begin{equation}
\label{ep5.2}
[a,a^+]=1
\end{equation}
Thus we have a representation for creation and annihilation operators
of the states of the string. The vacuum state annihilated 
by $z_{\mu}$ is the UET $\delta(z_{\mu})$,
and the orthonormalized states obtained by sucessive application of 
$\frac {d} {dz_{\mu}}$ to $\delta(z_{\mu})$ are:
\begin{equation}
\label{ep5.3}
F_n(z_{\mu})=\frac {{\delta}^{(n)}(z_{\mu})} {\sqrt{n!}}
\end{equation}

A general state of the string can be writen as:
\[\phi(x,\{z\})=[a_0(x)+a^{i_1}_{\mu_1}(x){\partial}^{\mu_1}_{i_1}+
a^{i_1 i_2}_{\mu_1\mu_2}(x){\partial}^{\mu_1}_{i_1}{\partial}^{\mu_2}_{i_2}
+...+...\]
\begin{equation}
\label{ep5.4}
+a^{i_1i_2...i_n}_{\mu_1\mu_2...\mu_n}(x){\partial}^{\mu_1}_{i_1}
{\partial}^{\mu_2}_{\i_2}...{\partial}^{\mu_n}_{i_n}+...+...]
\delta(\{z\})
\end{equation}
where $\{z\}$ denotes $(z_{1\mu},z_{2\mu},...,z_{n\mu},...,....)$, and 
$\phi$ is a UET of compact support in the set of variables $\{z\}$.
The functions
$a^{i_1i_2...i_n}_{\mu_1\mu_2...\mu_n}(x)$
are solutions of
\begin{equation}
\label{ep5.5}
\Box a^{i_1i_2...i_n}_{\mu_1\mu_2...\mu_n}(x)=0
\end{equation}

\subsection*{The case n$\rightarrow \infty$}

In this case
\begin{equation}
\label{ep5.6}
a=-z\;\;\;;\;\;\;a^+=-2z+\frac {d} {dz}
\end{equation}
we have
\begin{equation}
\label{ep5.7}
[a,a^+]=1
\end{equation}
The vacuum state annihilated by $a$ is $\delta(z)e^{z^2}$. The orthonormalized
states obtained by sucessive application of $a^+$ are:
\begin{equation}
\label{ep5.8}
{\hat{F}}_n(z)=2^{\frac {1} {4}}{\pi}^{\frac {1} {2}}
\frac {{\delta}^{(n)}(z)e^{z^2}} {\sqrt{n!}}
\end{equation}

\section{The Supersymmetric String}

\setcounter{equation}{0}

According to the treatment given in ref.\cite{tq1}  the action for the supersymmetric 
string is given by:
\begin{equation}
\label{ep6.1}
S=-\frac {1} {2\pi}\int\limits_{-\infty}^{\infty}
\int\limits_0^{\pi}
|{\dot{\Pi}}^2-{\Pi}^{'2}|\;d^2\sigma    
\end{equation}
where
\[{\dot{\Pi}}^{\mu}={\dot{X}}^{\mu}+\frac {i} {2}
\overline{\dot{\Theta}}
{\Gamma}^{\mu}\Theta-\frac {i} {2}\overline{\Theta}
{\Gamma}^{\mu}\dot{\Theta}\]
\begin{equation}
\label{ep6.2}
{\Pi}^{'\mu}=X^{'\mu}+\frac {i} {2}\overline{{\Theta}^{'}}
{\Gamma}^{\mu}\Theta-\frac {i} {2}\overline{\Theta}
{\Gamma}^{\mu}{\Theta}^{'}
\end{equation}
(see ref.\cite{tp13}) and $\Theta$ is a Dirac spinor.\\
We define
\begin{equation}
\label{ep6.3}
{\dot{X}}^{\mu}_{\infty}=\lim_{\tau\rightarrow\infty}{\dot{X}}^{\mu}(\tau,\sigma)
\end{equation}
Following ref.\cite{tq1} two possible
set of constraints for the string are: 
\begin{equation}
\label{ep6.4}
\begin{cases}
(\dot{\Pi}+\Pi^{'})^2=0\\
\Gamma\cdot{\dot{X}}_{\infty}=0
\end{cases}
\end{equation}
\begin{equation}
\label{ep6.5}
\begin{cases}
(\dot{\Pi}-\Pi^{'})^2=0\\
\Gamma\cdot{\dot{X}}_{\infty}=0
\end{cases}
\end{equation}
Thus, we have that to solve the system described by:
\begin{equation}
\label{ep6.6}
\begin{cases}
{\cal L}=|{\dot{\Pi}}^2-{\Pi^{'}}^2|\\
(\dot{\Pi}+\Pi^{'})^2=0\\
\Gamma\cdot{\dot{X}}_{\infty}=0\\
X_{\mu}(\tau,0)=X_{\mu}(\tau,\pi)=0\\
{\Theta}_{\mu}(\tau,0)={\Theta}_{\mu}(\tau,\pi)=0
\end{cases}
\end{equation}
or equivalently
\begin{equation}
\label{ep6.7}
\begin{cases}
{\cal L}=|{\dot{\Pi}}^2-{\Pi^{'}}^2|\\
(\dot{\Pi}-\Pi^{'})^2=0\\
\Gamma\cdot{\dot{X}}_{\infty}=0\\
X_{\mu}(\tau,0)=X_{\mu}(\tau,\pi)=0\\
{\Theta}_{\mu}(\tau,0)={\Theta}_{\mu}(\tau,\pi)=0
\end{cases}
\end{equation}
We define
\begin{equation}
\label{ep6.8}
{\cal L}_1={\dot{\Pi}}^2-{\Pi^{'}}^2
\end{equation}
Then the Euler-Lagrange equations for (\ref{ep6.6}),(\ref{ep6.7}) are
\[\frac {\partial} {\partial\tau}[Sgn({\cal L}_1)(
{\dot{X}}^{\mu}+\frac {i} {2}
\overline{\dot{\Theta}}
{\Gamma}^{\mu}\Theta-\frac {i} {2}\overline{\Theta}
{\Gamma}^{\mu}\dot{\Theta})]-\]
\begin{equation}
\label{ep6.9}
\frac {\partial} {\partial\sigma}
[Sgn({\cal L}_1)(
X^{'\mu}+\frac {i} {2}\overline{{\Theta}^{'}}
{\Gamma}^{\mu}\Theta-\frac {i} {2}\overline{\Theta}
{\Gamma}^{\mu}{\Theta}^{'})]=0
\end{equation}
\[\frac {\partial} {\partial\tau}[Sgn({\cal L}_1)(
{\dot{X}}^{\mu}+\frac {i} {2}
\overline{\dot{\Theta}}
{\Gamma}^{\mu}\Theta-\frac {i} {2}\overline{\Theta}
{\Gamma}^{\mu}\dot{\Theta})
({\Theta}^{+\beta}{\Omega}^{\mu}_{\beta\alpha})]-\]
\[\frac {\partial} {\partial\sigma}[Sgn({\cal L}_1)(
X^{'\mu}+\frac {i} {2}\overline{{\Theta}^{'}}
{\Gamma}^{\mu}\Theta-\frac {i} {2}\overline{\Theta}
{\Gamma}^{\mu}{\Theta}^{'})
({\Theta}^{+\beta}{\Omega}^{\mu}_{\beta\alpha})]+\]
\[Sgn({\cal L}_1)
({\dot{X}}^{\mu}+\frac {i} {2}
\overline{\dot{\Theta}}
{\Gamma}^{\mu}\Theta-\frac {i} {2}\overline{\Theta}
{\Gamma}^{\mu}\dot{\Theta})(
{\dot{\Theta}}^{+\beta}{\Omega}^{\mu}_{\beta\alpha})-\]
\begin{equation}
\label{ep6.10}
 Sgn({\cal L}_1)(X^{'\mu}+\frac {i} {2}\overline{{\Theta}^{'}}
{\Gamma}^{\mu}\Theta-\frac {i} {2}\overline{\Theta}
{\Gamma}^{\mu}{\Theta}^{'})(
{{\Theta}^{'}}^{+\beta}{\Omega}^{\mu}_{\beta\alpha})=0
\end{equation}
\[\frac {\partial} {\partial\tau}[Sgn({\cal L}_1)(
{\dot{X}}^{\mu}+\frac {i} {2}
\overline{\dot{\Theta}}
{\Gamma}^{\mu}\Theta-\frac {i} {2}\overline{\Theta}
{\Gamma}^{\mu}\dot{\Theta})
({\Omega}^{\mu}_{\alpha\beta}{\Theta}^{\beta})]-\]
\[\frac {\partial} {\partial\sigma}[Sgn({\cal L}_1)(
X^{'\mu}+\frac {i} {2}\overline{{\Theta}^{'}}
{\Gamma}^{\mu}\Theta-\frac {i} {2}\overline{\Theta}
{\Gamma}^{\mu}{\Theta}^{'})
({\Omega}^{\mu}_{\alpha\beta}{\Theta}^{\beta})]+\]
\[Sgn({\cal L}_1)
({\dot{X}}^{\mu}+\frac {i} {2}
\overline{\dot{\Theta}}
{\Gamma}^{\mu}\Theta-\frac {i} {2}\overline{\Theta}
{\Gamma}^{\mu}\dot{\Theta})(
{\Omega}^{\mu}_{\alpha\beta}{\dot{\Theta}}^{\beta})-\]
\begin{equation}
\label{ep6.11}
 Sgn({\cal L}_1)(X^{'\mu}+\frac {i} {2}\overline{{\Theta}^{'}}
{\Gamma}^{\mu}\Theta-\frac {i} {2}\overline{\Theta}
{\Gamma}^{\mu}{\Theta}^{'})(
{\Omega}^{\mu}_{\alpha\beta}{\Theta}^{'\beta})=0
\end{equation}
where ${\Omega}^{\mu}={\Gamma}^0{\Gamma}^{\mu}$

The solution for the equations (\ref{ep6.6}) are:
\begin{equation}
\label{ep6.12}
\begin{cases}
{\dot{X}}^{\mu}=\frac {i} {2}\overline{\Theta}
{\Gamma}^{\mu}\dot{\Theta}-
\frac {i} {2}
\overline{\dot{\Theta}}
{\Gamma}^{\mu}\Theta+{\dot{V}}^{\mu}\\
\dot{\Theta}+{\Theta}^{'}=0\\
{\dot{V}}^{\mu}+V^{'\mu}=0
\end{cases}
\end{equation}
and for (\ref{ep6.7}):
\begin{equation}
\label{ep6.13}
\begin{cases}
{\dot{X}}^{\mu}=\frac {i} {2}\overline{\Theta}
{\Gamma}^{\mu}\dot{\Theta}-
\frac {i} {2}
\overline{\dot{\Theta}}
{\Gamma}^{\mu}\Theta
+{\dot{V}}^{\mu}\\
\dot{\Theta}-{\Theta}^{'}=0\\
{\dot{V}}^{\mu}-V^{'\mu}=0
\end{cases}
\end{equation}
From (\ref{ep6.12}) we obtain
\begin{equation}
\label{ep6.14}
{\Theta}^{\alpha}=\sum\limits_{n=0}^{\infty}
c_n^{\alpha}e^{-2in(\tau-\sigma)}+
d_n^{+\alpha}e^{2in(\tau-\sigma)}
\end{equation}
\[V^{\mu}=x^{\mu}-\sum\limits_{n=1}^{\infty}2n(d_n^{\alpha}
{\Omega}^{\mu}_{\alpha\beta}d_n^{+\beta}-
c_n^{+\alpha}{\Omega}^{\mu}_{\alpha\beta}c_n^{\beta})\sigma +\]
\[(l^2p_{\mu}+\sum\limits_{n=1}^{\infty}2n(d_n^{\alpha}
{\Omega}^{\mu}_{\alpha\beta}d_n^{+\beta}-
c_n^{+\alpha}{\Omega}^{\mu}_{\alpha\beta}c_n^{\beta})\tau+\]
\begin{equation}
\label{ep6.15}
\frac {il} {2}
\sum\limits_{n=-\infty\;;\;n\neq 0}^{\infty}
\frac {a^{\mu}_n} {n} e^{-2in(\tau-\sigma)}
\end{equation}
\[X^{\mu}=x^{\mu}+l^2 p^{\mu}\tau+
\frac {il} {2}
\sum\limits_{s=-\infty\;;\;s\neq 0}^{\infty}
\frac {a^{\mu}_s} {s} e^{-2is(\tau-\sigma)}+\]
\[\sum\limits_{s=-\infty\;;\;s\neq 0}^{\infty}
\sum\limits_{n=0\;;\;n+s\geqq 0}^{\infty}
i\frac {2n+s} {2s}c_n^{+\alpha}{\Omega}^{\mu}_{\alpha\beta}
c_{n+s}^{\beta}e^{-2is(\tau-\sigma)}+\]
\[\sum\limits_{s=-\infty\;;\;s\neq 0}^{-1}
\sum\limits_{n=0\;;\;n+s\leqq -1}^{\infty}
i\frac {2n+s} {2s}c_n^{+\alpha}{\Omega}^{\mu}_{\alpha\beta}
d_{-n-s}^{+\beta}e^{-2is(\tau-\sigma)}-\]
\[\sum\limits_{s=1}^{\infty}
\sum\limits_{n=1\;;\;s-n\geqq 0}^{\infty}
i\frac {2n-s} {2s}d_n^{\alpha}{\Omega}^{\mu}_{\alpha\beta}
c_{s-n}^{\beta}e^{-2is(\tau-\sigma)}-\]
\begin{equation}
\label{ep6.16}
\sum\limits_{s=-\infty\;;\;s\neq 0}^{\infty}
\sum\limits_{n=1\;;\;n-s\geqq 1}^{\infty}
i\frac {2n-s} {2s}d_n^{\alpha}{\Omega}^{\mu}_{\alpha\beta}
d_{n-s}^{+\beta}e^{-2is(\tau-\sigma)}
\end{equation}
Using these solutions eq. (\ref{ep6.10}) and (\ref{ep6.11}) transforms into:
\begin{equation}
\label{ep6.17}
{\dot{\Theta}}^{+\beta}{\Omega}^{\mu}_{\beta\alpha}p_{\mu}=0
\end{equation}
\begin{equation}
\label{ep6.18}
p_{\mu}{\Omega}^{\mu}_{\alpha\beta}{\Theta}^{\beta}=0
\end{equation}
which are consistent  with the constraints for (\ref{ep6.6}), 
namely:
\begin{equation}
\label{ep6.19}
\begin{cases}
p^2|\Psi>=(\Gamma\cdot p)^2|\Psi>=0\\
\Gamma\cdot p|\Psi>=0
\end{cases}
\end{equation}
where $|\Psi>$ is the physical state of the string.
It is sufficient to solve  the second constraint 
because it implies the first one.

Similarly from (\ref{ep6.13}) we obtain
\begin{equation}
\label{ep6.20}
{\Theta}^{\alpha}=\sum\limits_{n=0}^{\infty}
c_n^{\alpha}e^{-2in(\tau+\sigma)}+
d_n^{+\alpha}e^{2in(\tau+\sigma)}
\end{equation}
\[V^{\mu}=x^{\mu}+\sum\limits_{n=1}^{\infty}2n(d_n^{\alpha}
{\Omega}^{\mu}_{\alpha\beta}d_n^{+\beta}-
c_n^{+\alpha}{\Omega}^{\mu}_{\alpha\beta}c_n^{\beta})\sigma +\]
\[(l^2p_{\mu}+\sum\limits_{n=1}^{\infty}2n(d_n^{\alpha}
{\Omega}^{\mu}_{\alpha\beta}d_n^{+\beta}-
c_n^{+\alpha}{\Omega}^{\mu}_{\alpha\beta}c_n^{\beta})\tau+\]
\begin{equation}
\label{ep6.21}
\frac {il} {2}
\sum\limits_{n=-\infty\;;\;n\neq 0}^{\infty}
\frac {a^{\mu}_n} {n} e^{-2in(\tau+\sigma)}
\end{equation}
\[X^{\mu}=x^{\mu}+l^2 p^{\mu}\tau+
\frac {il} {2}
\sum\limits_{s=-\infty\;;\;s\neq 0}^{\infty}
\frac {a^{\mu}_s} {s} e^{-2is(\tau+\sigma)}+\]
\[\sum\limits_{s=-\infty\;;\;s\neq 0}^{\infty}
\sum\limits_{n=0\;;\;n+s\geqq 0}^{\infty}
i\frac {2n+s} {2s}c_n^{+\alpha}{\Omega}^{\mu}_{\alpha\beta}
c_{n+s}^{\beta}e^{-2is(\tau+\sigma)}+\]
\[\sum\limits_{s=-\infty\;;\;s\neq 0}^{-1}
\sum\limits_{n=0\;;\;n+s\leqq -1}^{\infty}
i\frac {2n+s} {2s}c_n^{+\alpha}{\Omega}^{\mu}_{\alpha\beta}
d_{-n-s}^{+\beta}e^{-2is(\tau+\sigma)}-\]
\[\sum\limits_{s=1}^{\infty}
\sum\limits_{n=1\;;\;s-n\geqq 0}^{\infty}
i\frac {2n-s} {2s}d_n^{\alpha}{\Omega}^{\mu}_{\alpha\beta}
c_{s-n}^{\beta}e^{-2is(\tau+\sigma)}-\]
\begin{equation}
\label{ep6.22}
\sum\limits_{s=-\infty\;;\;s\neq 0}^{\infty}
\sum\limits_{n=1\;;\;n-s\geqq 1}^{\infty}
i\frac {2n-s} {2s}d_n^{\alpha}{\Omega}^{\mu}_{\alpha\beta}
d_{n-s}^{+\beta}e^{-2is(\tau+\sigma)}
\end{equation}
\begin{equation}
\label{ep6.23}
\Gamma\cdot p|\Psi>=0
\end{equation}

\section{A representation of  the states of the closed supersymmatric string}

\setcounter{equation}{0}

\subsection*{The case n finite}

As in ref.\cite{tq1}, for n finite we have:
\begin{equation}
\label{ep7.1}
\begin{cases}
a=-z\;\;\;;\;\;\;a^+=\frac {d} {dz}\\
c=\frac {d} {d\theta}\;\;\;;\;\;\;c^+=\theta\\
d=\frac {d} {d\vartheta}\;\;\;;\;\;\;d^+=\vartheta
\end{cases}
\end{equation}
\begin{equation}
\label{ep7.2}
[a,a^+]=\{c,c^+\}=\{d,d^+\}=1
\end{equation}
where $-z$ and $d/dz$ are operators over CUET and $\theta$ and
$\vartheta$ are Grassman variables with scalar product defined by:
\begin{equation}
\label{ep7.3}
<f,g>=\int f(\theta)e^{\theta{\theta}^+}g^+(\theta)\;d\theta\;
d{\theta}^+
\end{equation}
As for the bosonic string, a general state of the
supersymmetric string can be writen as: 
\[{\Psi}_{\alpha}(x,\{z\},\{\theta\},\{\vartheta\})=[c_0a_{\alpha 0}(x)+
c(1,0,0)a^{i_1}_{\alpha{\mu}_1}(x){\partial}^{{\mu}_1}_{i_1}+\]
\[c(0,1,0)a^{j_1}_{\alpha{\alpha}_1}(x){\theta}^{{\alpha}_1}_{j_1}+
c(0,0,1)a^{k_1}_{\alpha{\beta}_1}(x){\vartheta}^{{\beta}_1}_{k_1}+
\cdot\cdot\cdot+\]
\[c(m,n,p)a_{\alpha{\mu}_1\cdot\cdot\cdot{\mu}_m
{\alpha}_1\cdot\cdot\cdot{\alpha}_n
{\beta}_1\cdot\cdot\cdot{\beta}_p}^{i_1\cdot
\cdot\cdot i_m j_1\cdot\cdot\cdot j_n
k_1\cdot\cdot\cdot k_p}(x){\partial}^{{\mu}_1}_{i_1}
\cdot\cdot\cdot{\partial}^{{\mu}_m}_{i_m}
{\theta}^{{\alpha}_1}_{j_1}\cdot\cdot\cdot{\theta}^{{\alpha}_n}_{j_n}
{\vartheta}^{{\beta}_1}_{k_1}\cdot\cdot\cdot{\vartheta}^{{\beta}_p}_{k_p}+\]
\begin{equation}
\label{ep7.4}
+\cdot\cdot\cdot+\cdot\cdot\cdot]\delta(\{z\})
\end{equation}
where $c(m,n,p)$ are constants to be evaluated.
In this case the physical state $\Psi$ is a spinor whose components
are defined in (\ref{ep7.4}). 
\begin{equation}
\label{ep7.5}
{\Psi}(x,\{z\},\{\theta\},\{\vartheta\})=\left(
\begin{array}{l}
{\Psi}_1(x,\{z\},\{\theta\},\{\vartheta\})\\ 
{\Psi}_2(x,\{z\},\{\theta\},\{\vartheta\})\\ 
\;\;\;\;\;\;\;\;\;\;\;\;\;\cdot\\
\;\;\;\;\;\;\;\;\;\;\;\;\;\cdot\\
\;\;\;\;\;\;\;\;\;\;\;\;\;\cdot\\
{\Psi}_n(x,\{z\},\{\theta\},\{\vartheta\})
\end{array}
\right)
\end{equation}
\begin{equation}
\label{ep7.6}
a_{{\mu}_1\cdot\cdot\cdot{\mu}_m
{\alpha}_1\cdot\cdot\cdot{\alpha}_n
{\beta}_1\cdot\cdot\cdot{\beta}_p}^{i_1\cdot
\cdot\cdot i_m j_1\cdot\cdot\cdot j_n
k_1\cdot\cdot\cdot k_p}(x)=\left(
\begin{array}{l}
a_{1{\mu}_1\cdot\cdot\cdot{\mu}_m
{\alpha}_1\cdot\cdot\cdot{\alpha}_n
{\beta}_1\cdot\cdot\cdot{\beta}_p}^{i_1\cdot
\cdot\cdot i_m j_1\cdot\cdot\cdot j_n
k_1\cdot\cdot\cdot k_p}(x)\\
a_{2{\mu}_1\cdot\cdot\cdot{\mu}_m
{\alpha}_1\cdot\cdot\cdot{\alpha}_n
{\beta}_1\cdot\cdot\cdot{\beta}_p}^{i_1\cdot
\cdot\cdot i_m j_1\cdot\cdot\cdot j_n
k_1\cdot\cdot\cdot k_p}(x)\\
\;\;\;\;\;\;\;\;\;\;\;\;\;\;\;\;\;\cdot\\
\;\;\;\;\;\;\;\;\;\;\;\;\;\;\;\;\;\cdot\\
\;\;\;\;\;\;\;\;\;\;\;\;\;\;\;\;\;\cdot\\
a_{n{\mu}_1\cdot\cdot\cdot{\mu}_m
{\alpha}_1\cdot\cdot\cdot{\alpha}_n
{\beta}_1\cdot\cdot\cdot{\beta}_p}^{i_1\cdot
\cdot\cdot i_m j_1\cdot\cdot\cdot j_n
k_1\cdot\cdot\cdot k_p}(x)
\end{array}
\right)
\end{equation}
Its components are solutions of
\begin{equation}
\label{ep7.7}
{\Gamma}_{\mu}^{\beta\alpha}{\partial}^{\mu}
{\Phi}_{\alpha}(x,\{z\},\{\theta\},\{\vartheta\})=0
\end{equation}
\begin{equation}
\label{ep7.8}
{\Gamma}_{\mu}^{\beta\alpha}{\partial}^{\mu}
a_{\alpha{\mu}_1\cdot\cdot\cdot{\mu}_m
{\alpha}_1\cdot\cdot\cdot{\alpha}_n
{\beta}_1\cdot\cdot\cdot{\beta}_p}^{i_1\cdot
\cdot\cdot i_m j_1\cdot\cdot\cdot j_n
k_1\cdot\cdot\cdot k_p}(x)=0
\end{equation}

\subsection*{The case n$\rightarrow \infty$}

In this case:
\begin{equation}
\label{ep7.10}
\begin{cases}
a=-z\;\;\;;\;\;\;a^+=-2z+\frac {d} {dz}\\
c=\frac {d} {d\theta}\;\;\;;\;\;\;c^+=\theta\\
d=\frac {d} {d\vartheta}\;\;\;;\;\;\;d^+=\vartheta
\end{cases}
\end{equation}
\begin{equation}
\label{ep7.11}
[a,a^+]=\{c,c^+\}=\{d,d^+\}=1
\end{equation}
and the expression for the physical state of the string is
similar to the finite case.

\section{The Field of the Supersymmetric String }

\setcounter{equation}{0}

According to (\ref{ep6.17}) and section 7 the equation for the string field
is given by
\begin{equation}
\label{ep8.1}
(\Gamma\cdot\partial)\Psi(x,\{z\},\{\theta\},\{\vartheta\})=0
\end{equation}
where $\{z\}$ denotes $(z_{1\mu},z_{2\mu},...,z_{n\mu},...,....)$, and 
$\Psi$ is a CUET in the set of variables $\{z\}$.
Any UET of compact support can be writen as a development of
$\delta(\{z\})$ and its derivatives. Thus we have:
\[{\Psi}(x,\{z\},\{\theta\},\{\vartheta\})=[c_0A_0(x)+
c(1,0,0)A^{i_1}_{{\mu}_1}(x){\partial}^{{\mu}_1}_{i_1}+\]
\[c(0,1,0)A^{j_1}_{{\alpha}_1}(x){\theta}^{{\alpha}_1}_{j_1}+
c(0,0,1)A^{k_1}_{{\beta}_1}(x){\vartheta}^{{\beta}_1}_{k_1}+
\cdot\cdot\cdot+\]
\[c(m,n,p)A_{{\mu}_1\cdot\cdot\cdot{\mu}_m
{\alpha}_1\cdot\cdot\cdot{\alpha}_n
{\beta}_1\cdot\cdot\cdot{\beta}_p}^{i_1\cdot
\cdot\cdot i_m j_1\cdot\cdot\cdot j_n
k_1\cdot\cdot\cdot k_p}(x){\partial}^{{\mu}_1}_{i_1}
\cdot\cdot\cdot{\partial}^{{\mu}_m}_{i_m}
{\theta}^{{\alpha}_1}_{j_1}\cdot\cdot\cdot{\theta}^{{\alpha}_n}_{j_n}
{\vartheta}^{{\beta}_1}_{k_1}\cdot\cdot\cdot{\vartheta}^{{\beta}_p}_{k_p}+\]
\begin{equation}
\label{ep8.2}
+\cdot\cdot\cdot+\cdot\cdot\cdot]\delta(\{z\})
\end{equation}
where the quantum fields 
$A_{{\mu}_1\cdot\cdot\cdot{\mu}_m
{\alpha}_1\cdot\cdot\cdot{\alpha}_n
{\beta}_1\cdot\cdot\cdot{\beta}_p}^{i_1\cdot
\cdot\cdot i_m j_1\cdot\cdot\cdot j_n
k_1\cdot\cdot\cdot k_p}(x) $
are solutions of
\begin{equation}
\label{ep8.3}
(\Gamma\cdot\partial)
A_{{\mu}_1\cdot\cdot\cdot{\mu}_m
{\alpha}_1\cdot\cdot\cdot{\alpha}_n
{\beta}_1\cdot\cdot\cdot{\beta}_p}^{i_1\cdot
\cdot\cdot i_m j_1\cdot\cdot\cdot j_n
k_1\cdot\cdot\cdot k_p}(x)=0
\end{equation}
The propagator of the string field can be exppresed in terms of the propagators 
of the component fields:
\[{\Delta}_{\alpha\beta}(x-x^{'},\{z\},\{z^{'}\},\{\theta\},\{{\theta}^{'}\},
\{\vartheta\},\{{\vartheta}^{'}\})=[c_0^2
{\Delta}_{\alpha\beta}(x-x^{'})+\cdot\cdot\cdot+\]
\[c^2(m,n,p){\Delta}_{\alpha\beta{\mu}_1\cdot\cdot\cdot{\mu}_m
{\alpha}_1\cdot\cdot\cdot{\alpha}_n{\beta}_1\cdot\cdot\cdot{\beta}_p
{\nu}_1\cdot\cdot\cdot{\nu}_m{\gamma}_1\cdot\cdot\cdot{\gamma}_n
{\delta}_1\cdot\cdot\cdot{\delta}_p}^{i_1\cdot\cdot\cdot i_m
j_1\cdot\cdot\cdot j_n k_1\cdot\cdot\cdot k_p l_1\cdot\cdot\cdot l_m
s_1\cdot\cdot\cdot s_n t_1\cdot\cdot\cdot t_p}(x-x^{'})\]
\[{\partial}^{{\mu}_1}_{i_1}\cdot\cdot\cdot{\partial}^{{\mu}_m}_{i_m}
{\partial}^{'{\nu}_1}_{l_1}\cdot\cdot\cdot{\partial}^{'{\nu}_m}_{l_m}
{\theta}^{{\alpha}_1}_{j_1}\cdot\cdot\cdot{\theta}^{{\alpha}_n}_{j_n}
{\vartheta}^{{\beta}_1}_{k_1}\cdot\cdot\cdot{\vartheta}^{{\beta}_p}_{k_p}
{\theta}^{'+{\gamma}_1}_{s_1}\cdot\cdot\cdot
{\theta}^{'+{\gamma}_n}_{s_n}\]
\begin{equation}
\label{ep8.4}
{\vartheta}^{'+{\delta}_1}_{t_1}\cdot\cdot\cdots
{\vartheta}^{'+{\delta}_p}_{t_p}+\cdot\cdot\cdot]
\delta(\{z\},\{z^{'}\})
\end{equation}
Writing
\[A_{\alpha{\mu}_1\cdot\cdot\cdot{\mu}_m
{\alpha}_1\cdot\cdot\cdot{\alpha}_n
{\beta}_1\cdot\cdot\cdot{\beta}_p}^{i_1\cdot
\cdot\cdot i_m j_1\cdot\cdot\cdot j_n
k_1\cdot\cdot\cdot k_p}(x)=\int\limits_{-\infty}^{\infty}
a_{\alpha{\mu}_1\cdot\cdot\cdot{\mu}_m
{\alpha}_1\cdot\cdot\cdot{\alpha}_n
{\beta}_1\cdot\cdot\cdot{\beta}_p}^{i_1\cdot
\cdot\cdot i_m j_1\cdot\cdot\cdot j_n
k_1\cdot\cdot\cdot k_p}(k)e^{-ik_{\mu}x^{\mu}}+\]
\begin{equation}
\label{ep8.5}
b_{\alpha{\mu}_1\cdot\cdot\cdot{\mu}_m
{\alpha}_1\cdot\cdot\cdot{\alpha}_n
{\beta}_1\cdot\cdot\cdot{\beta}_p}^{+i_1\cdot
\cdot\cdot i_m j_1\cdot\cdot\cdot j_n
k_1\cdot\cdot\cdot k_p}(k)e^{ik_{\mu}x^{\mu}}
\;d^{\nu-1}k
\end{equation}
We may define the operators of annihilation and creation for a string as:
\[a_{\alpha}(k,\{z\},\{\theta\},\{\vartheta\})=
[c_0a_{0\alpha}(k)+c(1,0,0)a_{\alpha\mu_1}^{i_1}(k)
\partial_{i_1}^{\mu_1}+\]
\[c(0,1,0)a_{\alpha{\alpha}_1}^{j_1}(k){\theta}^{{\alpha}_1}_{j_1}+
c(0,0,1)a_{\alpha{\beta}_1}^{k_1}(k){\vartheta}^{{\beta}_1}_{k_1}
+\cdot\cdot\cdot+\]
\[c(m,n,p)a_{\alpha{\mu}_1\cdot\cdot\cdot{\mu}_m
{\alpha}_1\cdot\cdot\cdot{\alpha}_n
{\beta}_1\cdot\cdot\cdot{\beta}_p}^{i_1\cdot
\cdot\cdot i_m j_1\cdot\cdot\cdot j_n
k_1\cdot\cdot\cdot k_p}(k){\partial}^{{\mu}_1}_{i_1}
\cdot\cdot\cdot{\partial}^{{\mu}_m}_{i_m}
{\theta}^{{\alpha}_1}_{j_1}\cdot\cdot\cdot{\theta}^{{\alpha}_n}_{j_n}
{\vartheta}^{{\beta}_1}_{k_1}\cdot\cdot\cdot{\vartheta}^{{\beta}_p}_{k_p}+\]
\begin{equation}
\label{ep8.6}
+...+...]\delta(\{z\})
\end{equation}
\[a^+_{\alpha}(k,\{z\},\{\theta\},\{\vartheta\})=
[c_0a^+_{0\alpha}(k)+c(1,0,0)a_{\alpha\mu_1}^{+i_1}(k)
\partial_{i_1}^{\mu_1}+\]
\[c(0,1,0)a_{\alpha{\alpha}_1}^{+j_1}(k){\theta}^{+{\alpha}_1}_{j_1}+
c(0,0,1)a_{\alpha{\beta}_1}^{+k_1}(k){\vartheta}^{+{\beta}_1}_{k_1}
+\cdot\cdot\cdot+\]
\[c(m,n,p)a_{\alpha{\mu}_1\cdot\cdot\cdot{\mu}_m
{\alpha}_1\cdot\cdot\cdot{\alpha}_n
{\beta}_1\cdot\cdot\cdot{\beta}_p}^{+i_1\cdot
\cdot\cdot i_m j_1\cdot\cdot\cdot j_n
k_1\cdot\cdot\cdot k_p}(k){\partial}^{{\mu}_1}_{i_1}
\cdot\cdot\cdot{\partial}^{{\mu}_m}_{i_m}
{\theta}^{+{\alpha}_1}_{j_1}\cdot\cdot\cdot{\theta}^{+{\alpha}_n}_{j_n}
{\vartheta}^{+{\beta}_1}_{k_1}\cdot\cdot\cdot{\vartheta}^{+{\beta}_p}_{k_p}+\]
\begin{equation}
\label{ep8.7}
+...+...]\delta(\{z\})
\end{equation}
where the constans $c(m,n,p)$ are solution of:
\[c^{\ast}(m,n,p)
{\vartheta}^{+{\beta}_p}_{k_p}\cdot\cdot\cdot{\vartheta}^{+{\beta}_1}_{k_1}
{\theta}^{+{\alpha}_n}_{j_n}\cdot\cdot\cdot{\theta}^{+{\alpha}_1}_{j_1}
a_{\alpha{\mu}_1\cdot\cdot\cdot{\mu}_m
{\alpha}_1\cdot\cdot\cdot{\alpha}_n
{\beta}_1\cdot\cdot\cdot{\beta}_p}^{+i_1\cdot
\cdot\cdot i_m j_1\cdot\cdot\cdot j_n
k_1\cdot\cdot\cdot k_p}(k)=\]
\begin{equation}
\label{ep8.8}
c(m,n,p)a_{\alpha{\mu}_1\cdot\cdot\cdot{\mu}_m
{\alpha}_1\cdot\cdot\cdot{\alpha}_n
{\beta}_1\cdot\cdot\cdot{\beta}_p}^{+i_1\cdot
\cdot\cdot i_m j_1\cdot\cdot\cdot j_n
k_1\cdot\cdot\cdot k_p}(k)
{\theta}^{+{\alpha}_1}_{j_1}\cdot\cdot\cdot{\theta}^{+{\alpha}_n}_{j_n}
{\vartheta}^{+{\beta}_1}_{k_1}\cdot\cdot\cdot{\vartheta}^{+{\beta}_p}_{k_p}
\end{equation}
and define the creation and annihilation operators of the anti-string:
\[b^+_{\alpha}(k,\{z\},\{\theta\},\{\vartheta\})=
[c_0b^+_{0\alpha}(k)+c(1,0,0)b_{\alpha\mu_1}^{+i_1}(k)
\partial_{i_1}^{\mu_1}+\]
\[c(0,1,0)b_{\alpha{\alpha}_1}^{+j_1}(k){\theta}^{{\alpha}_1}_{j_1}+
c(0,0,1)b_{\alpha{\beta}_1}^{+k_1}(k){\vartheta}^{{\beta}_1}_{k_1}
+\cdot\cdot\cdot+\]
\[c(m,n,p)b_{\alpha{\mu}_1\cdot\cdot\cdot{\mu}_m
{\alpha}_1\cdot\cdot\cdot{\alpha}_n
{\beta}_1\cdot\cdot\cdot{\beta}_p}^{+i_1\cdot
\cdot\cdot i_m j_1\cdot\cdot\cdot j_n
k_1\cdot\cdot\cdot k_p}(k){\partial}^{{\mu}_1}_{i_1}
\cdot\cdot\cdot{\partial}^{{\mu}_m}_{i_m}
{\theta}^{{\alpha}_1}_{j_1}\cdot\cdot\cdot{\theta}^{{\alpha}_n}_{j_n}
{\vartheta}^{{\beta}_1}_{k_1}\cdot\cdot\cdot{\vartheta}^{{\beta}_p}_{k_p}+\]
\begin{equation}
\label{ep8.9}
+...+...]\delta(\{z\})
\end{equation}
\[b_{\alpha}(k,\{z\},\{\theta\},\{\vartheta\})=
[c_0b_{0\alpha}(k)+c(1,0,0)b_{\alpha\mu_1}^{i_1}(k)
\partial_{i_1}^{\mu_1}+\]
\[c(0,1,0)b_{\alpha{\alpha}_1}^{j_1}(k){\theta}^{+{\alpha}_1}_{j_1}+
c(0,0,1)b_{\alpha{\beta}_1}^{k_1}(k){\vartheta}^{+{\beta}_1}_{k_1}
+\cdot\cdot\cdot+\]
\[c(m,n,p)b_{\alpha{\mu}_1\cdot\cdot\cdot{\mu}_m
{\alpha}_1\cdot\cdot\cdot{\alpha}_n
{\beta}_1\cdot\cdot\cdot{\beta}_p}^{i_1\cdot
\cdot\cdot i_m j_1\cdot\cdot\cdot j_n
k_1\cdot\cdot\cdot k_p}(k){\partial}^{{\mu}_1}_{i_1}
\cdot\cdot\cdot{\partial}^{{\mu}_m}_{i_m}
{\theta}^{+{\alpha}_1}_{j_1}\cdot\cdot\cdot{\theta}^{+{\alpha}_n}_{j_n}
{\vartheta}^{+{\beta}_1}_{k_1}\cdot\cdot\cdot{\vartheta}^{+{\beta}_p}_{k_p}+\]
\begin{equation}
\label{ep8.10}
+...+...]\delta(\{z\})
\end{equation}
As a consecuence we have
\[\Psi_{\alpha}(x,\{z\},\{\theta\},\{\vartheta\})=
\int\limits_{-\infty}^{\infty}
a_{\alpha}(x,\{z\},\{\theta\},\{\vartheta\})
e^{-ik_{\mu}x^{\mu}}+\]
\begin{equation}
\label{ep8.11}
b^+_{\alpha}(x,\{z\},\{\theta\},\{\vartheta\})
e^{ik_{\mu}x^{\mu}}\;d^{\nu-1}x
\end{equation}
If we define
\begin{equation}
\label{ep8.12}
\begin{cases}
[\;\;\;,\;\;\;]_{n+p+1}=[\;\;\;,\;\;\;] ; \;\;\;n+p+1\;\;\; even\\
[\;\;\;,\;\;\;]_{n+p+1}=\{\;\;\;,\;\;\;\} ; \;\;\;n+p+1\;\;\; odd
\end{cases}
\end{equation}
with
\[[a_{\alpha{\mu}_1\cdot\cdot\cdot{\mu}_m
{\alpha}_1\cdot\cdot\cdot{\alpha}_n
{\beta}_1\cdot\cdot\cdot{\beta}_p}^{i_1\cdot
\cdot\cdot i_m j_1\cdot\cdot\cdot j_n
k_1\cdot\cdot\cdot k_p}(k),
a_{\beta{\nu}_1\cdot\cdot\cdot{\nu}_m
{\gamma}_1\cdot\cdot\cdot{\gamma}_n
{\delta}_1\cdot\cdot\cdot{\delta}_p}^{+l_1\cdot
\cdot\cdot l_m s_1\cdot\cdot\cdot s_n
t_1\cdot\cdot\cdot t_p}(k^{'})]_{n+p+1}=\]
\begin{equation}
\label{ep8.13}
f_{\alpha\beta{\mu}_1\cdot\cdot\cdot{\mu}_m
{\alpha}_1\cdot\cdot\cdot{\alpha}_n
{\beta}_1\cdot\cdot\cdot{\beta}_p
{\nu}_1\cdot\cdot\cdot{\nu}_m
{\gamma}_1\cdot\cdot\cdot{\gamma}_n
{\delta}_1\cdot\cdot\cdot{\delta}_p}^{
i_1\cdot\cdot\cdot i_m
j_1\cdot\cdot\cdot j_n
k_1\cdot\cdot\cdot k_p
l_1\cdot\cdot\cdot l_m
s_1\cdot\cdot\cdot s_n
t_1\cdot\cdot\cdot t_p}(k)\delta(k-k^{'})
\end{equation}
Then
\[\{a_{\alpha}(k,\{z\},\{\theta\},\{\vartheta\}),
a^+_{\beta}(k^{'},\{z{'}\},\{\theta^{'}\},\{\vartheta^{'}\})\}=
c_0^2f_{0\alpha\beta}(k)+\cdot\cdot\cdot+\]
\[c^2(m,n,p)f_{\alpha\beta{\mu}_1\cdot\cdot\cdot{\mu}_m
{\alpha}_1\cdot\cdot\cdot{\alpha}_n{\beta}_1\cdot\cdot\cdot{\beta}_p
{\nu}_1\cdot\cdot\cdot{\nu}_m{\gamma}_1\cdot\cdot\cdot{\gamma}_n
{\delta}_1\cdot\cdot\cdot{\delta}_p}^{i_1\cdot\cdot\cdot i_m
j_1\cdot\cdot\cdot j_n k_1\cdot\cdot\cdot k_p l_1\cdot\cdot\cdot l_m
s_1\cdot\cdot\cdot s_n t_1\cdot\cdot\cdot t_p}(k-k^{'})\]
\[{\partial}^{{\mu}_1}_{i_1}\cdot\cdot\cdot{\partial}^{{\mu}_m}_{i_m}
{\partial}^{'{\nu}_1}_{l_1}\cdot\cdot\cdot{\partial}^{'{\nu}_m}_{l_m}
{\theta}^{{\alpha}_1}_{j_1}\cdot\cdot\cdot{\theta}^{{\alpha}_n}_{j_n}
{\vartheta}^{{\beta}_1}_{k_1}\cdot\cdot\cdot{\vartheta}^{{\beta}_p}_{k_p}
{\theta}^{'+{\gamma}_1}_{s_1}\cdot\cdot\cdot
{\theta}^{'+{\gamma}_n}_{s_n}\]
\begin{equation}
\label{ep8.14}
{\vartheta}^{'+{\delta}_1}_{t_1}\cdot\cdot\cdots
{\vartheta}^{'+{\delta}_p}_{t_p}+\cdot\cdot\cdot]
\delta(\{z\},\{z^{'}\})
\end{equation}
and for the anti-string
\[[b_{\alpha{\mu}_1\cdot\cdot\cdot{\mu}_m
{\alpha}_1\cdot\cdot\cdot{\alpha}_n
{\beta}_1\cdot\cdot\cdot{\beta}_p}^{i_1\cdot
\cdot\cdot i_m j_1\cdot\cdot\cdot j_n
k_1\cdot\cdot\cdot k_p}(k),
b_{\beta{\nu}_1\cdot\cdot\cdot{\nu}_m
{\gamma}_1\cdot\cdot\cdot{\gamma}_n
{\delta}_1\cdot\cdot\cdot{\delta}_p}^{+l_1\cdot
\cdot\cdot l_m s_1\cdot\cdot\cdot s_n
t_1\cdot\cdot\cdot t_p}(k^{'})]_{n+p+1}=\]
\begin{equation}
\label{ep8.15}
g_{\alpha\beta{\mu}_1\cdot\cdot\cdot{\mu}_m
{\alpha}_1\cdot\cdot\cdot{\alpha}_n
{\beta}_1\cdot\cdot\cdot{\beta}_p
{\nu}_1\cdot\cdot\cdot{\nu}_m
{\gamma}_1\cdot\cdot\cdot{\gamma}_n
{\delta}_1\cdot\cdot\cdot{\delta}_p}^{
i_1\cdot\cdot\cdot i_m
j_1\cdot\cdot\cdot j_n
k_1\cdot\cdot\cdot k_p
l_1\cdot\cdot\cdot l_m
s_1\cdot\cdot\cdot s_n
t_1\cdot\cdot\cdot t_p}(k)\delta(k-k^{'})
\end{equation}
Thus
\[\{b_{\alpha}(k,\{z\},\{\theta\},\{\vartheta\}),
b^+_{\beta}(k^{'},\{z{'}\},\{\theta^{'}\},\{\vartheta^{'}\})\}=
c_0^2g_{0\alpha\beta}(k)+\cdot\cdot\cdot+\]
\[c^2(m,n,p)g_{\alpha\beta{\mu}_1\cdot\cdot\cdot{\mu}_m
{\alpha}_1\cdot\cdot\cdot{\alpha}_n{\beta}_1\cdot\cdot\cdot{\beta}_p
{\nu}_1\cdot\cdot\cdot{\nu}_m{\gamma}_1\cdot\cdot\cdot{\gamma}_n
{\delta}_1\cdot\cdot\cdot{\delta}_p}^{i_1\cdot\cdot\cdot i_m
j_1\cdot\cdot\cdot j_n k_1\cdot\cdot\cdot k_p l_1\cdot\cdot\cdot l_m
s_1\cdot\cdot\cdot s_n t_1\cdot\cdot\cdot t_p}(k)\delta(k-k^{'})\]
\[{\partial}^{{\mu}_1}_{i_1}\cdot\cdot\cdot{\partial}^{{\mu}_m}_{i_m}
{\partial}^{'{\nu}_1}_{l_1}\cdot\cdot\cdot{\partial}^{'{\nu}_m}_{l_m}
{\theta}^{+{\alpha}_1}_{j_1}\cdot\cdot\cdot{\theta}^{+{\alpha}_n}_{j_n}
{\vartheta}^{+{\beta}_1}_{k_1}\cdot\cdot\cdot{\vartheta}^{+{\beta}_p}_{k_p}
{\theta}^{'{\gamma}_1}_{s_1}\cdot\cdot\cdot
{\theta}^{'{\gamma}_n}_{s_n}\]
\begin{equation}
\label{ep8.16}
{\vartheta}^{'{\delta}_1}_{t_1}\cdot\cdot\cdots
{\vartheta}^{'{\delta}_p}_{t_p}+\cdot\cdot\cdot]
\delta(\{z\},\{z^{'}\})
\end{equation}

\section{The Action for the Field of the Supersymmetric
String}

\setcounter{equation}{0}

\subsection*{The case n finite}

The action for the free supersymmetric closed string field is:
\[S_{free}=\iiiint
\oint\limits_{\{\Gamma_1\}}\oint\limits_{\{\Gamma_2\}}
 \int\limits_{-\infty}^{\infty}
i\overline{\Psi}(x,\{z_1\},\{{\theta}_1\},\{{\vartheta}_1\})e^{\{z_1\}{\cdot}\{z_2\}}
e^{\{{\theta}^+_1\}{\cdot}\{{\theta}_1\}}e^{\{{\vartheta}^+_1\}
{\cdot}\{{\vartheta}_1\}}\]
\begin{equation}
\label{ep9.1}
\not\!{\partial}\Psi(x,\{z_2\},\{{\theta}_1\},\{{\vartheta}_1\})
\;d^{\nu}x\;\{dz_1\}\{dz_2\}\{d{\theta}^+_1\}\{d\theta_1\}\{d{\vartheta}^+_1\}
\{d\vartheta_1\}
\end{equation}
where $\not\!\partial=\Gamma\cdot\partial$\\
A possible interaction is given by:
\[S_{int}=\lambda\;\iiiint\iiiint\oint\limits_{\{\Gamma_1\}}
\oint\limits_{\{\Gamma_2\}}
\oint\limits_{\{\Gamma_3\}}
\oint\limits_{\{\Gamma_4\}}
\int\limits_{-\infty}^{\infty}
\overline{\Psi}(x,\{z_1\},\{{\theta}_1\},\{{\vartheta}_1\})
e^{\{z_1\}{\cdot}\{z_2\}}
e^{\{{\theta}^+_1\}{\cdot}\{{\theta}_1\}}
e^{\{{\vartheta}^+_1\}{\cdot}\{{\vartheta}_1\}}\]
\[\Psi(x,\{z_2\},\{{\theta}_1\},\{{\vartheta}_1\})
e^{\{z_2\}{\cdot}\{z_3\}}
e^{\{{\theta}_1\}{\cdot}\{{\theta}^+_2\}}
e^{\{{\vartheta}_1\}{\cdot}\{{\vartheta}^+_2\}}
\]
\[\overline{\Psi}(x,\{z_3\},\{{\theta}_2\},\{{\vartheta}_2\})
e^{\{z_3\}{\cdot}\{z_4\}}
e^{\{{\theta}^+_2\}{\cdot}\{{\theta}_2\}}
e^{\{{\vartheta}^+_2\}{\cdot}\{{\vartheta}_2\}}\]
\[\Psi(x,\{z_4\},\{{\theta}_2\},\{{\vartheta}_2\})\;
d^{\nu}x\;\{dz_1\}\{dz_2\}\{dz_3\}\{dz_4\}\]
\begin{equation}
\label{ep9.2}
\{d{\theta}^+_1\}\{d\theta_1\}\{d{\vartheta}^+_1\}
\{d\vartheta_1\}
\{d{\theta}^+_2\}\{d\theta_2\}\{d{\vartheta}^+_2\}
\{d\vartheta_2\}
\end{equation}
Both, $S_{free}$ and $S_{int}$ are non-local as expected.

\subsection*{The case n$\rightarrow \infty$}

In this case:
\[S_{free}=\iiiint
\oint\limits_{\{\Gamma_1\}}\oint\limits_{\{\Gamma_2\}}
 \int\limits_{-\infty}^{\infty}
i\overline{\Psi}(x,\{z_1\},\{{\theta}_1\},\{{\vartheta}_1\})e^{\{z_1\}{\cdot}\{z_2\}}
e^{\{{\theta}^+_1\}{\cdot}\{{\theta}_1\}}e^{\{{\vartheta}^+_1\}
{\cdot}\{{\vartheta}_1\}}\]
\begin{equation}
\label{ep9.3}
\not\!{\partial}\Psi(x,\{z_2\},\{{\theta}_1\},\{{\vartheta}_1\})
\;d^{\nu}x\;\{d{\eta}_1\}\{d{\eta}_2\}
\{d{\theta}^+_1\}\{d\theta_1\}\{d{\vartheta}^+_1\}
\{d\vartheta_1\}
\end{equation}
where
\begin{equation}
\label{ep9.4}
d\eta(z)=\frac {e^{-z^2}} {\sqrt{2}\;\pi}
\end{equation}
and
\[S_{int}=\lambda\;\iiiint\iiiint\oint\limits_{\{\Gamma_1\}}
\oint\limits_{\{\Gamma_2\}}
\oint\limits_{\{\Gamma_3\}}
\oint\limits_{\{\Gamma_4\}}
\int\limits_{-\infty}^{\infty}
\overline{\Psi}(x,\{z_1\},\{{\theta}_1\},\{{\vartheta}_1\})
e^{\{z_1\}{\cdot}\{z_2\}}
e^{\{{\theta}^+_1\}{\cdot}\{{\theta}_1\}}
e^{\{{\vartheta}^+_1\}{\cdot}\{{\vartheta}_1\}}\]
\[\Psi(x,\{z_2\},\{{\theta}_1\},\{{\vartheta}_1\})
e^{\{z_2\}{\cdot}\{z_3\}}
e^{\{{\theta}_1\}{\cdot}\{{\theta}^+_2\}}
e^{\{{\vartheta}_1\}{\cdot}\{{\vartheta}^+_2\}}\]
\[\overline{\Psi}(x,\{z_3\},\{{\theta}_2\},\{{\vartheta}_2\})
e^{\{z_3\}{\cdot}\{z_4\}}
e^{\{{\theta}^+_2\}{\cdot}\{{\theta}_2\}}
e^{\{{\vartheta}^+_2\}{\cdot}\{{\vartheta}_2\}}\]
\[\Psi(x,\{z_4\},\{{\theta}_2\},\{{\vartheta}_2\})\;
d^{\nu}x\;\{d{\eta}_1\}\{d{\eta}_2\}\{d{\eta}_3\}\{d{\eta}_4\}\]
\begin{equation}
\label{ep9.5}
\{d{\theta}^+_1\}\{d\theta_1\}\{d{\vartheta}^+_1\}
\{d\vartheta_1\}
\{d{\theta}^+_2\}\{d\theta_2\}\{d{\vartheta}^+_2\}
\{d\vartheta_2\}
\end{equation}
The convolution
of two propagators of  the string field is:
\begin{equation}
\label{ep9.6}
\hat{{\Delta}}_{\alpha\beta}(k,\{z_1\},\{z_2\},\{\theta_1\},\{{\theta}^{'}_1\},
\{\vartheta_1\},\{{\vartheta}^{'}_1\})\ast 
\hat{{\Delta}}_{\alpha\beta}(k^{'},\{z_3\},\{z_4\},\{\theta_2\},\{{\theta}^{'}_2\},
\{\vartheta_2\},\{{\vartheta}^{'}_2\})
\end{equation}
where $\ast$ denotes the convolution
of Ultradistributions of Exponential Type  
on the $k$ variable only.
With the use of the result
\begin{equation}
\label{ep9.7}
\frac {1} {\rho}\ast\frac {1} {\rho}=-\pi^2\ln\rho
\end{equation}
$(\rho=x_1^2+x_2^2+\cdot\cdot\cdot+x_{\nu}^2)$ 
in euclidean space
and
\begin{equation}
\label{ep9.8}
\frac {1} {\rho\pm i0}\ast\frac {1} {\rho\pm i0}=
 \mp i\pi^2\ln(\rho\pm i0)
\end{equation}
($\rho=x_0^2-x_1^2-\cdot\cdot\cdot-x_{\nu-1}^2)$ 
in minkowskian space,
the convolution of two string field propagators is finite.

\section{Discussion}

We have decided to begin this paper, for the benefit
of the reader, with a summary of the main characteristics
of Ultradistributions of Exponential Type and their Fourier
transform.

We have shown that UET are appropriate for
the description 
in a consistent way superstring and superstring field theories.
By means of  a Lagrangian for the superstring 
we have obtained the non-linear Euler-Lagrange equations and 
solve them.
We have given the movement equation for the 
field of the superstring and solve it with the use of CUET.
We have shown that this superstring field 
is a linear superposition of CUET.
We have evaluated the propagator for the superstring field,
and calculate the convolution of two of them, taking
into account that superstring field theory is a non-local theory
of  UET of an infinite number of complex variables,
For practical calculations and experimental results 
we have given expressions that
involve only a finite number of variables.

As a final remark we would like to point out that our formulae
for convolutions follow from  general definitions. They are not
regularized expresions

\end{document}